\documentclass[]{article}

\usepackage{graphicx}

\begin{document}

\begin{center}
{\Large Random loose packing and an order parameter for the parking lot model}\\
\vspace{5mm}
K. Hern\'andez and L.I. Reyes\\
Departamento de F\'{\i}sica, Universidad Sim\'on Bol\'{\i}var, 
Apartado 89000, Caracas 1080-A, Venezuela
\end{center}

\begin{abstract} 

We have obtained the random loose packing fraction of the parking lot model (PLM) by taking the limit
of infinite compactivity in the two-variable statistical description of Tarjus and Viot
for the PLM. The PLM is a stochastic model of adsorption and desorption 
of particles on a substrate that have been used as a model for compaction of granular materials. 
An order parameter $\rho$ is introduced 
to characterize how far from a steady state situation the model is. Thus, configurations with $\rho<1$ age. 
We propose that $\rho$ can be a starting point in order to stablish a connection between Edwards' statistical mechanics
and granular hydrodynamics.

\end{abstract}

\section{Introduction}

More than twenty years ago Edwards and Oakeshott proposed a statistical
mechanics framework for granular materials in mechanical equilibrium \cite{Edwards}. The idea
was to replace the energy by the total volume of the sample $V$. Thus the entropy 
of the system is defined as $S=\log\Omega$, where $\Omega$ is the number of
stable states for a given volume $V$ (and given number of grains $N$). 
The quantity equivalent to temperature within this description is the compactivity 
$\chi=(\partial S/\partial V)^{-1}$. In connection with Edwards proposal, 
Aste {\it et al} has found for static packings of spheres an invariant distribution 
of Voronoi volumes at the grain
level \cite{Aste}. Also, the random close packing fraction $\phi_{rcp}$ and the random loose
packing fraction $\phi_{rlp}$ have been associated with configurations with $\chi\rightarrow 0$
and $\chi\rightarrow\infty$, respectively \cite{advances,nature,veryloose}.

It has been shown recently that two granular samples with the same packing fraction
may not have the same properties, and an aditional macroscopic variable must be 
introduced in the statistical description. In reference \cite{luis} was found that 
we can produce two samples with the same $\phi$ but different stresses.
Indeed, the stress tensor $\sigma$ has been included in a more general statistical description \cite{Ed2005}
(see \cite{luis} and references therein).

We deal in this article with the so called {\it parking lot model} (PLM) introduced by Nowak {\it et al}
in the context of experiments of compaction of grains \cite{pre98}.
Starting with the statistical mechanics for the PLM proposed by Tarjus and Viot \cite{TV},
we obtain the random loose packing fraction for this model and introduce an order parameter 
$\rho$ that characterizes how far from a steady state situation the model is. 
We propose that a quantity analog to $\rho$ can be used as the order parameter
in the continuum description of slow and dense granular flows by Aranson and Tsimring,
mediating how solid and fluid is the form of the stress tensor \cite{aranson2001,aranson2003}.

The PLM is a model of random adsorption and desorption of particles on a substrate. 
Particles are disorbed with rate $p_-$ and adsorbed with rate
$p_+$, with a no overlapping condition. For a given initial condition of the substrate,
the model converges to a stationary state packing fraction $\phi_e$ around which it fluctuates.
$\phi_e$ depends only on the ratio $K=p_+/p_-$, which allow us to map the parameter $K$ to $\Gamma$ or
$Q$ of references \cite{pre98,pre2005}. For large $K$ we have 
$\phi_e=1-1/\log K$ \cite{krap} and the convergence to a stationary state is very slow, 
reminiscent of glassy behavior \cite{glassy}. The PLM is though to represent an average column of grains. 

Tarjus and Viot characterized the configurations produced by the PLM with two variables \cite{TV}. One of them
is the packing fraction $\phi$ and the other is the insertion probability $\Phi$, which is the available line
fraction for a new insertion. Tarjus and Viot recognized the need for an additional variable because of
some memory effects observed in experiments that can be reproduced within the PLM if we change $K$ in
the course of a Monte Carlo simulation: we can generate two configurations with
the same $\phi$ but with different subsequent evolution and 
by allowing a variation in $K$, for a given finite time, we can obtain more dense substrates \cite{talbot}. 
Thus, the history of the configuration is encoded in $\Phi$, which is a structure variable \cite{euro}.
The aditional variable $\Phi$ is only needed to characterize configurations which
are not produced in steady state, since in steady state the insertion probability $\Phi_e$ is given by 
(eq. 2 in \cite{talbot}):
\begin{equation}\label{Phie}
\Phi_e=(1-\phi)\exp[-\phi/(1-\phi)].
\end{equation}

\begin{figure}
\begin{center}\includegraphics[scale=0.35]{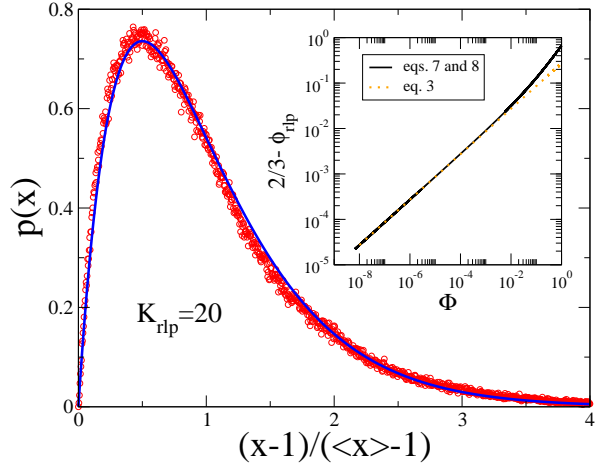}\end{center}
\caption{ Distribution of Voronoi lenghts of the PLM for $K_{rlp}\approx 20$ obtained with Monte Carlo Simulations (circles). 
The continuum line is a gamma distribution
with shape parameter $k=2$ \cite{Aste,us}. 
In the inset we show $2/3-\phi_{rlp}$ vs. the insertion probability $\Phi$ for the PLM as given by eqs. \ref{rlp1} and \ref{rlp2}. 
The dotted line is equation \ref{rlpPhi}. 
At $\Phi\approx 0.01$ the difference in $\phi_{rlp}$ is less than $1$\% .\label{phirlp}}
\end{figure}

\section{Results}

If $A$ is the total lenght available for a new insertion ($A=\Phi L$, where $L$ is the size of the system), 
for given $N$, $L$ and $A$ we have a configurational integral in terms of the gaps $h_i$ (eq. 25 in \cite{TV}):

\begin{equation}\label{Zgrande}
Z=\int_0^L\ldots\int_0^L dh_i^N \delta\left(L-N-\sum_{i=1}^N h_i\right)\delta\left( A-\sum_{i=1}^N \theta(h_i-1)(h_i-1)\right)
\end{equation}
with $\theta$ being the step function. Equation \ref{Zgrande} can be solved with a saddle point method 
in the limit of large $N$, $L$ and $A$, with $\phi=N/L$ and $\Phi=A/L$ fixed. 

In steady state, for $K=100$ we have $\Phi\approx 0.01$ \cite{talbot}. In the limit of small $\Phi$ we obtain a
random loose packing fraction by taking the limit $\chi\rightarrow\infty$ in the above description (see Appendix):
\begin{equation}\label{rlpPhi}
\left(\frac{2}{3}-\phi_{rlp}\right)\approx \sqrt{\frac{2}{27} }\Phi^{1/2}.
\end{equation}

Since the random loose packing of granular media depends on friction \cite{rlpSwinney}, from equation \ref{rlpPhi} 
we see that greater $\Phi$ suggest
grains with larger friction coefficient. For a given $\phi$, a greater $\Phi$ can also be 
associated to greater heterogeneity of voids at
the grain level. In the limit $\Phi\rightarrow 0$, from equation \ref{rlpPhi} we get $\phi_{rlp}=2/3$. This value for
$\phi_{rlp}$ correspond to smooth grains. A gamma distribution of Voronoi lenghts \cite{us} is obtained
for $K_{rlp}\approx 20$ (see figure \ref{phirlp}). 
The invariant distribution found in experiments by Aste {\it et al} for spheres
packings in mechanical equilibrium is also a gamma distribution \cite{Aste}.  

It have been reported that at a packing fraction near the $\phi_{rlp}$ found by us the process of compaction is slowered. 
Before reaching $\phi_e$, typically we have four differents regimes as we record the evolution of the
packing fraction in a Monte Carlo simulation of the PLM \cite{GG}. During a first stage, $\phi$ ``... increases rapidly until a
value of around $0.65$'' \cite{GG}, and from this point afterwards the increase in $\phi$ is considerably slower. Thus,
$\phi_{rlp}$ can play a role in the onset of jamming in granular materials, as have been speculated in reference \cite{GS}.

\begin{figure}
\begin{center}\includegraphics[scale=0.35]{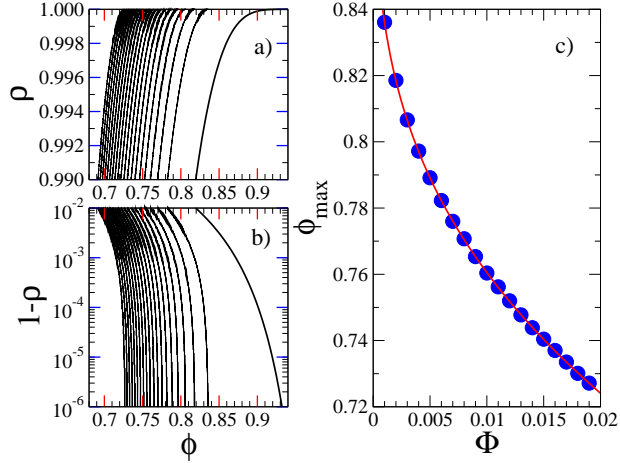}\end{center}
\caption{ a) Parameter $\rho$ (eq. \ref{rho}) as a function of packing fraction $\phi$ 
for different values of the insertion probability $\Phi$. 
From right
to left: $\Phi=0,0.001,0.002,...,0.018, 0.019$.
b) 1-$\rho$ on a logarithmic scale as a function of $\phi$ for different values of $\Phi$ (same data than in a)). 
From the intersection
with the horizontal axis we can estimate the maximum value of packing fraction $\phi_{max}$ for a given $\Phi$.  
c) Circles: $\phi_{max}$ vs. $\Phi$ as obtained from b). When $\phi\rightarrow\phi_{max}$, then $\rho\rightarrow 1$.
The solid line is equation \ref{Phie}, the steady state relation between $\phi$ and $\Phi$. \label{parOrden}}
\end{figure}

We introduce now a parameter $\rho$ which is, basically, the quotient between $Z$ given by equation \ref{Zgrande} 
and $Z_t$ which 
is the configurational integral without the restriction of having a definite value of $A$, {\it i.e.} 
when only the first $\delta$ in equation \ref{Zgrande} is considered. This leads to:
\begin{equation}\label{rho}
\rho(\phi,\Phi)=\exp(s-s_t),
\end{equation}
with $s_t=\phi+\phi\ln\left(\frac{1-\phi}{\phi} \right)$ \cite{TV} 
and the entropy density $s$ is given by equation \ref{entropy} in the Appendix. 
In figure \ref{parOrden}a we can see $\rho(\phi,\Phi)$.
For a given value of $\Phi$, $\rho \rightarrow 1$ as $\phi$ is increased up to a maximum value  $\phi_{max}$. 
This maximum value of $\phi$ can be estimated from a graph like the one shown in figure \ref{parOrden}b.

In figure \ref{parOrden}c it can be seen that $\rho\rightarrow 1$ is equivalent to say that we are 
approaching a steady state
situation. With this in mind, in figure \ref{orden} we plot the insertion probability $\Phi$ vs. 
the packing fraction $\phi$ 
for a given value of the parameter $\rho$. 
We did this by solving numerically the relevant equations needed for to evaluate equation \ref{rho}.
It is worth to remember that a $(\phi,\Phi)$ statistical description for the PLM makes sense only if we need to consider
configurations out of steady state, since in steady state these two variables are related by equation \ref{Phie}. 
Thus, configurations with $\rho<1$ age.

\section{Discussion}

\begin{figure}
\begin{center} \includegraphics[scale=0.08]{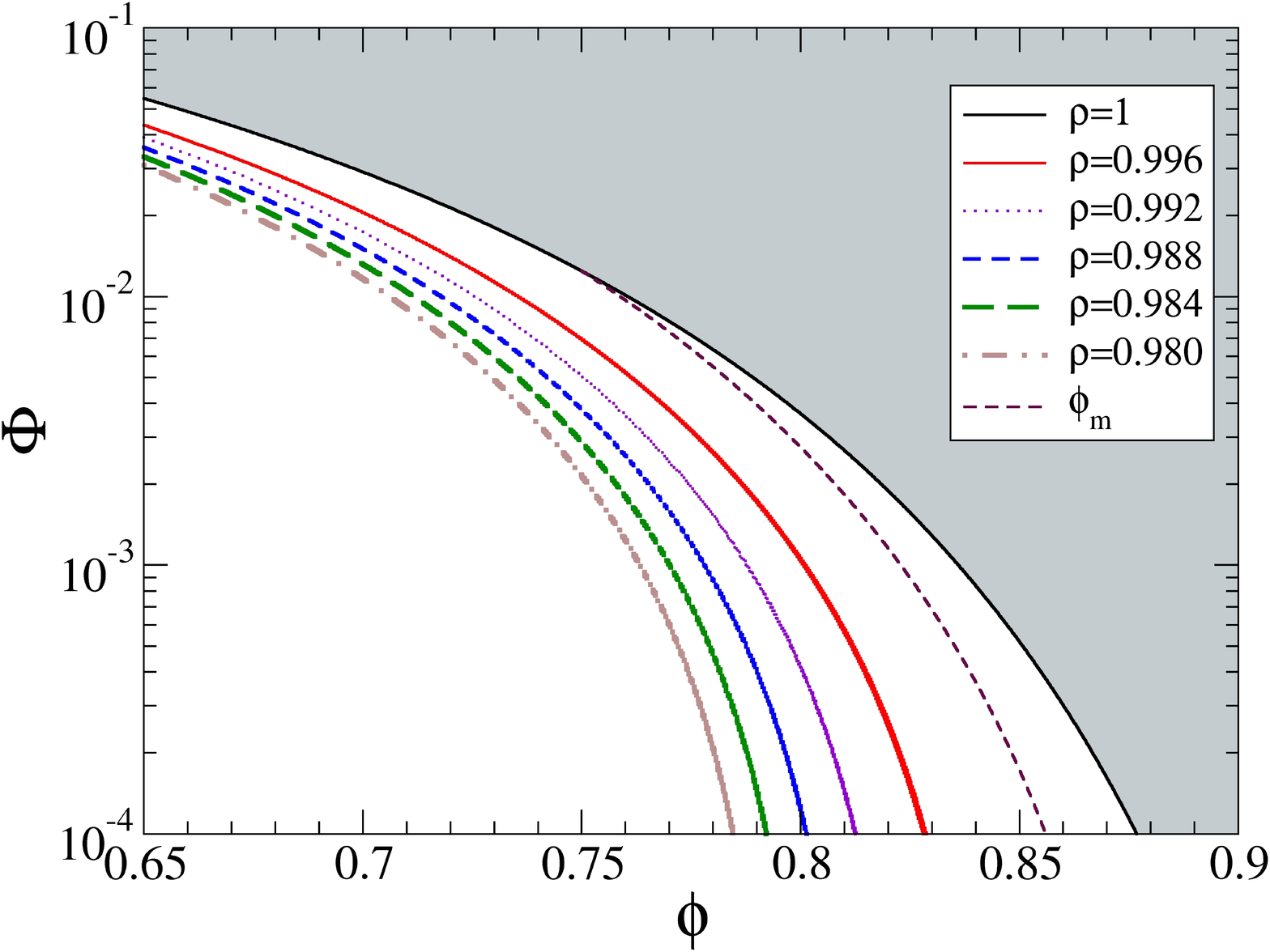} \end{center}
\caption[]{Insertion probability $\Phi$ vs packing fraction $\phi$ for different values of the parameter $\rho$. 
The particular case 
$\rho=1$ is given by equation \ref{Phie}, the steady state relation between $\phi$ and $\Phi$. 
The dark region are not available ($\phi,\Phi$) configurations and its frontier is defined by
the $\rho=1$ curve. Configurations with $\rho<1$ age.
The dotted line is a schematic representation of $\phi_m(K)$ \cite{talbot}.
\label{orden}}
\end{figure}

In the last stages of a Monte Carlo simulation of the PLM $\phi$ is increasing into $\phi_e$ with a small variation
in the insertion probability $\Phi$.  
Talbot, Tarjus and Viot found in reference \cite{talbot} that for $K>K_c\approx100$, there is a minimum in 
$\Phi$ as a function of time that occurs at a packing fraction $\phi_m(K)$.
This implies that for slow compaction if $K>K_c$ the system can increase its packing fraction
while increasing or decreasing $\Phi$, depending if $\phi$ is greater or smaller than $\phi_m$.
This should have consequences for {\it processes} (curves) in a $(\phi,\Phi)$ plane 
like the one shown in figure \ref{orden}. Thus, for $\phi>\phi_{rlp}$ in the PLM we have two zones in which we
expect different behaviour. In terms of $K$ these zones are: $K_{rlp}<K<K_c$ and $K>K_c$, with $K_{rlp}\approx 20$
and $K_c\approx 100$.

In reference \cite{trans} the authors reported a phase transition
when inserting slowly a rod into a column of grains: the system's response to shear changes at 
a certain packing fraction $\phi_c$.
$\phi_c$ can be localized by monitoring the change
in height $\Delta h$ of the column, after removing the rod, as a function of $\phi$ \cite{trans}. 
It would be interesting to put
the results of reference \cite{trans} in terms of a process that starts from a packing fraction $\phi_i$ on the
$\rho=1$ curve of figure \ref{orden} and ends on a packing fraction $\phi_f$, with $\phi_i>\phi_f$ \cite{trans}.

Can this two-variable description of the PLM be related to the volume-stress proposal of Edwards and others?
We believe that a connection can be made by using a quantity analog to $\rho(\phi,\Phi)$ given by equation \ref{rho} 
as the order parameter in the stress tensor $\sigma_{ij}$
proposed by Aranson and Tsimring in their continuum description of slow and dense granular flows 
\cite{aranson2001,aranson2003}:
\begin{equation}
\sigma_{ij}=\mu\left(\frac{\partial v_i}{\partial x_j}+\frac{\partial v_j}{\partial x_i}\right)+\sigma_{ij}^0 [\rho+(1-\rho)\delta_{ij}],
\end{equation} 
where $\sigma_{ij}^0$ is the stress under static conditions with the same geometry. Thus, 
$\rho$ in eq. \ref{rho} can control how fluid and solid is the form for
the stress tensor. For $\rho<1$ we have a partially fluidized granular medium. 
At best, this is a first step, a suggestion, towards a real connection between Edwards' statistical mechanics
and granular hydrodynamics \cite{rmpAranson}.

Finally, since configurations with $\rho<1$ in the PLM age we must say something about the relevant time scales for
this model. Kolan, Nowak and Tkachenko have found that the low relaxation frequency $\omega_L$
and the high relaxation frequency $\omega_H$ for this model are given by
$\omega_L=\frac{2p_+}{\delta_e}\exp(-2/\delta_e)$ and $\omega_H=p_+\delta$, 
with $\delta=(1-\phi)/\phi$ \cite{glassy}. Thus, in order to speak of thermodynamic processes in figure \ref{orden},
the observation time $t_{obs}$ \cite{Ma} must satisfy $1/\omega_H\ll t_{obs}\ll 1/\omega_L $. From figure 5 of
reference \cite{glassy}, we can see that this condition on $t_{obs}$ can be satisfied only for high packing fractions.
Only for $K>K_c$ we have at least two orders of magnitude of separation between $\omega_L$ and $\omega_H$.

\section{Conclusion}

We have obtained in this article the random loose packing fraction $\phi_{rlp}$ for the parking lot model, where
$\phi_{rlp}$ is the lower packing fraction in which we can find a sample in mechanical equilibrium.
We have done this by taking the limit of infinite compactivity in the statistical description of Tarjus and Viot,
in which a macro state is characterized by its packing fraction $\phi$ and its insertion probability $\Phi$.
The compactivity $\chi$ is the analog of temperature in the statistical mechanics for granular materials
proposed by Edwards. We have proposed an order parameter $\rho(\phi,\Phi)$ that characterizes how far from
a steady situation the model is. Thus, configurations with $\rho<1$ age. 
With $\rho$, we proposed a connection of statistical mechanics with the
continuum description of slow and dense granular flows by Aranson and Tsimring.
By considering the relevant time scales
for this model obtained by Kolan, Nowak and Tkachenko, we have argued that 
for blocked configurations ($\phi>\phi_{rlp}$), only for even higher
packing fractions we expect to be able to speak of thermodynamic processes for this model.

\section{Acknowledgments}

We thank Gustavo Guti\'errez for useful suggestions.
This work has been done under the PCP-FONACIT Franco-Venezuelan Program
{\it Dynamics and Statics of Granular Materials}, and was supported in part by DID of the Universidad Sim\'on Bol\'{\i}var.

\begin{figure}
\begin{center}\includegraphics[scale=0.35]{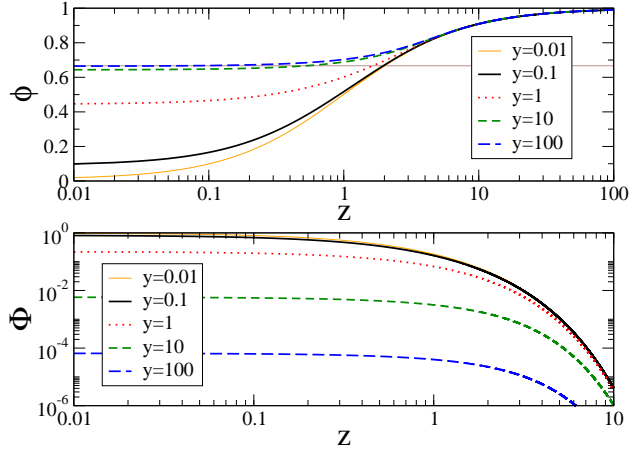}\end{center}
\caption{Equations 29 and 30 of reference \cite{TV}. For a given insertion probability $\Phi$, we have a maximum $z$, 
which implies a maximum
packing fraction $\phi_{max}$. The horizontal line corresponds to $\phi=2/3$, 
the maximum random loose packing fraction for
the PLM (see eq. \ref{rlpPhi}). \label{eqs2930}}
\end{figure}

\section{Appendix}

Tarjus and Viot obtained the entropy density $s(\phi,\Phi)$ ($Z=\exp(Ls)$) (eq. 28 in \cite{TV}):
\begin{equation}\label{entropy}
s=(1-\phi)z+y\Phi+\phi\ln\left(\frac{z+y[1-\exp(-z)]}{z(z+y)}\right)
\end{equation}
with $z=z(\phi,\Phi)$ and $y=y(\phi,\Phi)$ being solutions to the coupled equations 29 and 30 
of reference \cite{TV}, 
which can be seen in figure \ref{eqs2930}. 
We have that $z=(\partial S/\partial L)_{N,A}$ can be interpreted as the inverse of compactivity 
and $y=(\partial S/\partial A)_{N,L}$.

We can obtain the random loose packing fraction of this model by considering the limit $z\rightarrow 0$ 
($\chi\rightarrow\infty$) \cite{advances,nature,veryloose} in equations 29 and 30 of reference \cite{TV}. We obtain:

\begin{equation}\label{rlp1}
\Phi=\frac{1}{(1+y)\left [1+y+\frac{y^2}{2(1+y)}\right ]}
\end{equation}
and
\begin{equation}\label{rlp2}
\frac{1}{\phi_{rlp}}=1+\frac{1}{y}+\frac{y}{2(1+y)}
\end{equation}
from which we can eliminate $y$ to obtain $\phi_{rlp}(\Phi)$, as can be seen in the inset of figure \ref{phirlp}.

\end{document}